\documentstyle[12pt,fleqn,epsfig]{article}
\begin{document}

\begin{center}
{\bf INSTITUT~F\"{U}R~KERNPHYSIK,~UNIVERSIT\"{A}T~FRANKFURT}\\
D - 60486 Frankfurt, August--Euler--Strasse 6, Germany
\end{center}

\hfill IKF--HENPG/1--99

\vspace{0.5cm}

\vspace{0.5cm}
\begin{center}
   {\Large \bf On the Measurement of $D$--meson Yield}
\end{center}
\begin{center}
    {\Large \bf  in Nucleus--Nucleus 
     Collisions at the CERN SPS}  
\end{center}

\vspace{1cm}

\begin{center}
Marek Ga\'zdzicki\footnote{E--mail: marek@ikf.uni--frankfurt.de}\\
\vspace{0.3cm}
Institut f\"ur Kernphysik, Universit\"at Frankfurt \\
August--Euler--Strasse 6, D - 60486 Frankfurt, Germany\\[0.8cm]
\end{center}
\begin{center}
Christina Markert\footnote{E--mail: c.markert@gsi.de}\\
\vspace{0.3cm}
Gesellschaft f\"ur Schwerionenforschung (GSI) \\
Planckstr. 1, D - 64291 Darmstadt, Germany\\[0.8cm]
\end{center}

\begin{abstract}
\noindent
We argue that
the measurement of open charm gives a unique opportunity
to test the validity  of pQCD--based and statistical 
models of nucleus--nucleus collisions at high energies.
We show that various approaches used to estimate
$D$--meson multiplicity in central Pb+Pb collisions at
158 A$\cdot$GeV give predictions which differ by more than
a factor of 100.
Finally we demonstrate that decisive experimental results
concerning the open charm yield in A+A collisions can be obtained
using data  of the NA49 experiment at the CERN SPS. 
\end{abstract}

\vspace{1cm}
\begin{center}
{\it Talk given at NA49 Collaboration Meeting, March 8--13, 1999,
CERN}
\end{center}

\newpage

\section{Introduction}

The measurement of production of $D$--mesons in nucleus--nucleus
collisions at CERN SPS energies is a challenging experimental
adventure.
This is due to the short life time and expected low multiplicity 
of $D$--mesons which cause  the small experimental
signal originating  from $D$ decays to be  hidden in the large 
background composed of combinations of 'non--signal' tracks.

There is, however,  increasing  interest
in results on charm production in nucleus--nucleus (A+A)
collisions which
motivates various experimental groups to design
upgrades of the existing experiments or to build  new
experiments, which should allow to measure open charm 
production in A+A collisions \cite{add2,add3,NA50}.

The aim of this paper is to demonstrate that the first measurement
of $D$--meson production may be even possible using the current
set--up of the NA49 experiment \cite{NA49}. 

The paper starts with a summary of physics motivations
for charm measurement in A+A collisions at the CERN SPS (Section 2).
The expected multiplicity of $D$--mesons  in central Pb+Pb collisions
at 158 A$\cdot$GeV is discussed in Section 3.
An estimate of the 
statistical resolution of $D$--meson  measurement which can
be achieved using data of the NA49 experiment is given in Section 4.
Summary and conclusions (Section 5)  close the paper.

\newpage

\section{Physics Beyond QGP: pQCD vs Thermodynamics }

The main motivation for a broad experimental program
in which nucleus--nucleus collisions at high energies
are studied is the search for the Quark Gluon Plasma 
\cite{Co:75,Sh:80}.
An impressive set of data was collected during the last decades
and many unexpected phenomena were discovered
\cite{QM97}.
However, the question whether a transient QGP state is
created in the early stage of the collisions is still
under vivid discussion.
In our opinion this situation is due to the fact
that there is no consensus on which models of
high energy hadronic and nuclear collisions  should be
used to interpret experimental results.

The data are in qualitative and quantitative agreement
with the hypothesis of  QGP creation in A+A collisions
at the CERN SPS if a statistical model of the early stage
is used for their interpretation \cite{Ga:98}.
This model is, however, considered a non--orthodox
approach to hadronic and nuclear collisions as its basic
assumptions can not be derived from the commonly accepted
theory of strong interaction, QCD.
Its validity is, therefore,  more 
surprising than the conclusions reached within this model
concerning QGP creation at the CERN SPS.

On the other hand it is difficult to use QCD for the interpretation
of the experimental results. Problems arise because
almost all effects intuitively expected in the case of
a transition to QGP are in the domain of so--called
soft processes in which experimentally testable strict
predictions of QCD are not possible.
Attempts to build phenomenological
QCD--inspired models of soft processes 
are not very succesful \cite{Od:98}.
Conclusive interpretation of the data within these models
seems to be impossible as one can not estimate the uncertainties
introduced by the used approximations.

As a possible solution one tries to identify phenomena
sensitive to the  early stage in which
so--called hard QCD processes are believed to prevail \cite{Sa:97}.
In this case one hopes to obtain testable QCD predictions
using perturbative methods.
A well known example is the analysis of $J/\psi$
production in A+A collisions performed under the assumption
that the initial production of $c\overline{c}$ pairs is
described by pQCD.
However the limits of the applicability of pQCD are 
theoretically not well defined. 
Therefore any assumption 
based on validity 
of pQCD calculations and used to interpret the data
should be tested experimentally.
In the case of the $J/\psi$ example this test can be done
by the measurement of the open charm yield and its comparison to
pQCD predictions.

In connection with the above discussion of QCD--based models
it is interesting to consider a hypothetical model in which,
due to technical problems, strict experimentally testable  
predictions can not be obtained.
The model gives approximate predictions, however, the error
made due to the approximations is difficult to estimate.
Experimental results which are in agreement with the predictions
of the model can be treated as a `proof' of its validity.
On the other hand any disagreement between the model and the data
can be interpreted as due to the used approximations.
Therefore the model can not be falsified
\cite{Po:59} before a substantial improvment in its predictive power.
This example illustrates the logical problem
which one may encounter discussing
the validity of QCD--based models.

In the summarized exciting situation
concerning our understanding of strong interactions
the role of the experimental results on open charm production
in A+A collisions is unique.
\begin{itemize}
\item
Data on open charm should allow to test limits of applicability
of the perturbative QCD methods.\\
It  is often assumed that the charm quark is 
heavy enough for pQCD treatment \cite{Sa:95}.
Based on this hypothesis results on $J/\psi$ production
in A+A collisions are interpreted in the so--called
'suppression' models
\cite{Ge:98, Sa:97},
in which the suppression is calculated relative to the $J/\psi$
yield expected from the pQCD factorization theorem \cite{Co:98}
and data on $J/\psi$ production in p+p interactions.
It was, however, recently shown that the $J/\psi$ multiplicity
is, in good approximation, proportional to pion multiplicity \cite{Ga:98a}.
Thus the $J/\psi$  yield shows an $A$--dependence which is characteristic
of  soft QCD  processes.
This experimental observation motivates the question whether
pQCD can be used as a model of charm quark production
in hadronic and nuclear collisions.  
\item Data on open charm should allow to test the validity
of a non--orthodox statistical model of the early stage
of A+A collisions \cite{Ga:98}. \\
In this model charm quark multiplicity is large enough
to justify the use of the thermodynamical approximation.
Consequently not only the absolute yield of open charm but 
also its $A$--dependence are very different for the statistical
and the pQCD--based models. 
\end{itemize}

\newpage

\section{ Multiplicity of D--Mesons in Nuclear Collisions}

This section starts with a review of the data on $D^0$ and 
$\overline{D}^0$ production in proton--nucleon interactions,
which leads to an estimate of the mean multiplicity of
$D^0 + \overline{D}^0$ mesons, $\langle D^0 + \overline{D}^0 \rangle$,
in nucleon--nucleon, N+N, interactions at 158 A$\cdot$GeV.
This result is further used to predict the multiplicity of
$D^0 + \overline{D}^0$ mesons in central Pb+Pb collisions
at 158 A$\cdot$GeV using four different approaches.

\subsection{$\langle D^0 + \overline{D}^0 \rangle$ in N+N interactions}

There are six measurements of $D$--meson production cross section
at various collision energies 
ranging from 200 GeV to 800 GeV, which allow to estimate 
the mean multiplicity
of $D^0 + \overline{D}^0$ in p+N interactions 
\cite{Be:88, Al:96, Ag:84, Ag:88, Am:88, Le:94}.
In p+p interactions $\sigma(D^0 + \overline{D}^0)$ is measured at
360 GeV \cite{Ag:84}, 400 GeV \cite{Ag:88} and at 800 GeV \cite{Am:88}.
At 200 GeV \cite{Be:88}, 250 GeV \cite{Al:96} and 800 GeV \cite{Le:94}
the cross section in p+N interactions is estimated based on 
measured  cross section for p+A interactions assuming
that $\sigma(D^0 + \overline{D}^0)$ is proportional to $A$.
We note here that this assumption is not fully justified by the
experimental results \cite{Ga:98a} and  may lead to an additional
systematic error in estimating  $\sigma(D^0 + \overline{D}^0)$
in p+N interactions. 
In addition a symmetry of the  $x_F$ distribution of $D^0 + \overline{D}^0$
mesons
with respect to  $x_F = 0$ is assumed for the calculation of the integrated
cross section using measurements at  $x_F > 0$.
In the case of the  measurement at 200 GeV \cite{Be:88} only results 
for the sum of 
all
$D$--mesons are published; in this case we assume that 50\% of
them  are $D^0 + \overline{D}^0$ mesons.

The mean multiplicity of $D^0 + \overline{D}^0$ mesons in p+p
and p+N interactions is calculated as a ratio 
$\sigma(D^0 + \overline{D}^0)/\sigma$, where $\sigma$
is the inelastic cross section for p+p interaction at the
corresponding energy.
The values of $\langle D^0 + \overline{D}^0 \rangle_{NN}$ are given in Table 1
and plotted as a function of $\sqrt{s}$ in Fig. 1.
The results at $\sqrt{s} = 19.4$ GeV and  $\sqrt{s} = 21.7$ GeV
differ by a factor of about 8, which suggests a possible large
systematic uncertainty of the measurements at low energy not
accounted for in the quoted errors.
Taking this into account we estimate the mean multiplicity 
of  $D^0 + \overline{D}^0$ at $\sqrt{s} = 17.3$ GeV ($E_{LAB} = 158$ GeV)
in nucleon--nucleon interactions
by an arithmetic average of the measurements at $\sqrt{s} = 19.4$ GeV
and $\sqrt{s} = 21.7$ GeV, which gives:
\begin{equation}
\langle D^0 + \overline{D}^0 \rangle_{NN} = 2.0 \cdot 10^{-4}.
\end{equation}
A large systematic error up to several times the estimated value
should be kept in mind.

\subsection{$\langle D^0 + \overline{D}^0 \rangle$ in Central Pb+Pb Collisions
at 158 A$\cdot$GeV}

We discuss here four approaches which allow to estimate the mean
multiplicity of  $D^0 + \overline{D}^0$ in central Pb+Pb collisions,
$\langle D^0 + \overline{D}^0 \rangle_{PbPb}$, at 158 A$\cdot$GeV.
It is assumed that the central collisions are selected by
accepting only events with a low number of spectator nucleons
from the projectile nucleus and that  typical trigger conditions
of the NA49 experiment are used \cite{NA49}.
This results in a mean number of participant nucleons,
$\langle N_P \rangle$, 
of
about 350 \cite{PRL}.

\vspace{0.2cm}
\noindent
{\bf A}. It was recently observed \cite{Ga:98a}
that the mean multiplicity of $J/\psi$
mesons increases proportinally to the mean multiplicity of
negatively charged hadrons (more than 90\% $\pi^-$ mesons)  
when p+p, p+A and A+A collisions are considered (see Fig. 2).
It may be therefore assumed that a similar dependence on the
size of the colliding objects is also valid for $D$--mesons.
Following this assumption a mean multiplicity of
$D^0 + \overline{D}^0$ mesons can be calculated as:
\begin{equation}
\langle D^0 + \overline{D}^0 \rangle_{PbPb} = 
\langle D^0 + \overline{D}^0 \rangle_{NN} \cdot 
\frac {\langle h^- \rangle_{PbPb}}
{\langle h^- \rangle_{NN}} = 2 \cdot 10^{-4} \cdot
 \frac {700} {3.1} \approx 4.5 \cdot 10^{-2},
\end{equation}
where the values of $\langle h^- \rangle_{PbPb}$ and $\langle h^- \rangle_{NN}$
are taken from Ref. \cite{PRL} and Ref. \cite{Ga:95}, respectively.

\vspace{0.2cm}
\noindent
{\bf B}. It is usually assumed that due to the large mass of the charm quark,
production of charm can be calculated using perturbative QCD methods
\cite{Sa:95}.
This assumption leads to  the expectation that the cross section for charm
production increases as $A^2$ for all inelastic A+A collisions and 
that the mean multiplicity of open charm increases as
$\langle N_P \rangle^{4/3}$  for central A+A collisions.
Thus for central Pb+Pb collisions we get:
\begin{equation}
\langle D^0 + \overline{D}^0 \rangle_{PbPb} =
\langle D^0 + \overline{D}^0 \rangle_{NN} \cdot (\langle N_P \rangle/2)^{4/3}
\approx 2 \cdot 10^{-1}.
\end{equation}
This estimate agrees with a previously published prediction
for charm production based on  pQCD inspired models \cite{Br:98}. 

\vspace{0.2cm}
\noindent
{\bf C}. The  NA50 Collaboration found recently \cite{Sc:98}
that a model based on pQCD can not describe the dimuon invariant mass
spectrum between the $\phi$ and $J/\psi$ peaks in central Pb+Pb
collisions.
The spectrum, however, can be reproduced when the contribution from 
open charm decays is scaled up by a factor of about 3. 
Based on this analysis one may expect that:
\begin{equation}
\langle D^0 + \overline{D}^0 \rangle_{PbPb}
\approx 6 \cdot 10^{-1},
\end{equation}
i.e. it is equal to the multiplicity calculated in {\bf B}
multiplied by a factor of 3. 

\vspace{0.2cm}
\noindent
{\bf D}. The production of entropy and strangeness in A+A collisions
at the CERN SPS can be described by a statistical model  which assumes 
the creation of a Quark Gluon Plasma in the early stage of the 
collision \cite{Ga:98}. 
This fact and the observation that the $J/\psi$ yield is 
proportional to the pion yield triggered the hypothesis
that also charm production can be described by the
same statistical approach.
It was calculated, within this model \cite{Ga:98}, 
that the mean  number of $c$ and $\overline{c}$
quarks produced in central Pb+Pb collision at 158 A$\cdot$GeV
is about 17.
Based on p+p data \cite{Ag:88}
we assume here that about one third of them hadronize
as $D^0$ and  $\overline{D}^0$ mesons which gives:
\begin{equation}
\langle D^0 + \overline{D}^0 \rangle_{PbPb} 
\approx 6.
\end{equation}

\vspace{0.5cm}
\noindent
We summarize this section by observing that predictions
of the mean multiplicity of   $D^0 + \overline{D}^0$ mesons
in central Pb+Pb collisions at 158 A$\cdot$GeV
range from 4.5$\cdot$10$^{-2}$ to about 6 showing more than 
a factor 100 difference between minimum and maximum values.

\newpage

\section{$D$--Meson Measurement in NA49}

We consider here the possibility of a measurement of  $D^0$  
and $\overline{D}^0$ production in A+A collisions 
at 158 A$\cdot$GeV using the current set--up of the NA49
experiment at the CERN SPS \cite{NA49}.
We start from an introductory discussion on 
the means to distinguish $D^0$ decays
from combinatorial background and on the calculation of
the statistical error on $D^0$ multiplicity.
We continue with a brief summary of the relevant  properties
of the NA49 experiment and finally we present 
results of a simulation which yields an estimate of
the statistical resolution   
of  $D^0 + \overline{D}^0$ measurement 
in central Pb+Pb collisions
using the NA49 data.

Our analysis is based on the study of the two body
decay channels of  $D^0$ and $\overline{D}^0$ mesons:
\begin{equation}
D^0 \rightarrow K^- + \pi^+ {\rm ~~~~~and~~~~~}
\overline{D}^0 \rightarrow K^+ + \pi^- 
\end{equation}
for which the branching ratio is measured to be 
(3.85 $\pm$ 0.09)\% \cite{PDG}.
The mass and the proper life time
of  $D^0$ and $\overline{D}^0$ mesons are
$m_{D^0}$ = (1864.6 $\pm$ 0.5) MeV/c$^2$ and
$c \tau$ = 0.01244 cm, respectively \cite{PDG}. 

\subsection{Introductory Remarks}

Let us consider central Pb+Pb collision at CERN SPS energy
in which among thousands of produced particles  are also $D^0$
mesons.
The $D^0$ mesons decay
after a typical flight distance of about 0.1 cm 
($\approx \gamma \cdot c \tau$)   from the
collision point.
There are many decay channels possible, but for the reasons
discussed at the end of this paper we consider only decays of
a single type, $D^0 \rightarrow K^- + \pi^+$,
which happen only in 3.85 \% of all cases of $D^0$ decays.

For  simplicity of the initial  discussion we assume
that an ideal  detector is placed around the collision 
point, 
i.e. for all charged particles electric charge, mass and momentum vector
at a given reference plane are measured.

For all ($K^-$, $\pi^+$) pairs originating from $D^0$ decays
(signal pairs) the following conditions have to be
approximately (within experimental resolution) fulfilled:
\begin{enumerate}
\item
the trajectories of $K^-$ and $\pi^+$ intersect in a point
(decay point) which is different from the Pb+Pb collision point,
\item
the distribution of the $D^0$ life time in its own c.m.s. frame is an
exponential distribution with a mean value equal to
$c \tau = 0.0124$ cm,
\item
energy and momentum conservation laws are fulfilled at the decay
point when $D^0$ decay hypothesis is assumed,
\item
the angular distribution of the decay products in the $D^0$ c.m.s. is
isotropic.
\end{enumerate}

Conditions 1--4 can be fulfilled 
(within experimental resolution) by a set of ($K^-$, $\pi^+$) pairs
which do not originate from the same $D^0$ decay (background pairs)
only by chance.
The probability of this happening decreases with increasing experimental
resolution.

As a useful example we consider the condition 3 of energy momentum conservation
at the decay vertex.
For practical reasons one often quantifies deviations from
energy momentum conservation, assuming the $D^0$ decay
hypothesis, 
by calculating
the invariant mass, $M(K^-, \pi^+)$, for a pair
of negatively and positively charged particles and
comparing it to the known mass of the $D^0$ meson.
Almost all signal pairs are expected to be distributed in a narrow
interval $\Delta M$ centered around $m_{D^0}$.
The $M$ distribution for background pairs
is significantly broader and it has no characteristic peak
structure at  $m_{D^0}$.
The smearing of the $M$ distribution for signal pairs results
from the finite resolution of momentum measurement.
With increasing resolution the size of the interval $\Delta M$
in which almost all signal pairs are included  decreases 
consequently resulting in lower number of accepted background pairs.

In the high resolution limit almost only signal pairs
(and almost no background pairs)
are selected when taking pairs from $\Delta M$.
In this case the  total number of signal pairs, $N_S$, in $N_{EV}$
events can be approximated by the  measured number of pairs in the
$\Delta M$ interval, $N_{\Delta M}$.
An estimate of  the mean multiplicity of signal pairs 
is  therefore given by
\begin{equation}
\langle N^E_S \rangle = \frac { N_{\Delta M} } { N_{EV} } 
\end{equation}
and its statistical error can be calculated as
\begin{equation}
\sigma (\langle N^E_S \rangle) = \frac 
{ \sqrt{\langle N_{\Delta M} \rangle} } { \sqrt{N_{EV}} } \approx
\frac
{ \sqrt{\langle N^E_S \rangle} } { \sqrt{N_{EV}} } 
\end{equation}
assuming a Poissonian distribution of pair  
multiplicity in the interval $\Delta M$. 

In the limit of poor resolution the number of background pairs,  $N_B$,
in the interval $\Delta M$ is much larger than the number of
signal pairs.
Therefore 
in order to estimate the number of signal pairs, $N^E_S$, an
estimate of the number of background pairs, $N^E_B$ is needed:
\begin{equation}
N^E_S = N_{\Delta M} - N^E_B 
\end{equation}
The estimate of $N^E_B$ is model dependent and therefore
it has a systematic error which however will not be discussed here.
Various models can be used.
As an example let us mention a 'mixed event' model of the background
in which background pairs are constructed by selecting 
particles from different events.
Independent of the model used it is reasonable to assume
that the statistical error of $N^E_B$ can be made much smaller
than the statistical error of $N_{\Delta M}$.
Thus the statistical error of $\langle N^E_S \rangle$ can
be calculated as
\begin{equation}
\sigma(\langle N^E_S \rangle) = \frac { \sqrt{\langle N_{\Delta M} \rangle} }
{ \sqrt{N_{EV} } } \approx 
\frac { \sqrt{\langle N^E_B \rangle} } { \sqrt{N_{EV} } },
\end{equation}
assuming Poissonian distribution of pair multiplicity in the
interval $\Delta M$.

We observe that in the limit of high resolution the statistical
error of the signal  is almost fully defined by the number of
signal pairs, whereas in the limit of poor resolution it is defined
by the number  of background pairs.

One can define the statistical significance of the measurement 
as the ratio $\langle N^E_S \rangle/\sigma(\langle N^E_S \rangle$
which in the case of the poor resolution limit is given by
\begin{equation}
\frac {\langle N^E_S \rangle} {\sigma(\langle N^E_S \rangle)} =
\frac {\langle N^E_S \rangle} {\sqrt{\langle N^E_B \rangle}}
\cdot \sqrt{ N_{EV} }.
\end{equation}
Thus in order to maximize the statistical significance of the result
one should select the acceptance (in the example above 
the size of the  $\Delta M$ interval)  such that the ratio
$\langle N^E_S \rangle/\sqrt{\langle N^E_B \rangle}$
reaches a maximum.

\subsection{The NA49 Experiment}

The NA49 experiment \cite{NA49} at the CERN SPS was designed
and constructed to search for  signals of the Quark Gluon Plasma
created at the early stage of nucleus--nucleus collisions.
Basic detectors of the NA49 set--up are four Time Projection 
Chambers (TPC's), which allow for a precise tracking
of charged particles.
Two medium size TPC's of 3 m$^3$ gas volume each are located
inside of two magnets with up to 1.5 T field strength each.
Two large size TPC's of 20 m$^3$ gas volume each are positioned
downstream of the magnets for high precision energy loss measurement
and acceptance coverage in the forward direction.
Overall the TPC's acceptance coverage amounts to up to 80\%
of all produced charged particles (this  
number depends on the reaction studied and magnetic field --
target configuration).

The TPC's measure up to 234 space points on tracks of up
to 13 m length.
This allows  a precise determination of sign of electric
charge, momentum vector ($\sigma(p)/p^2 \approx 10^{-4}$ (GeV/c)$^{-1}$)
and energy loss ($ \approx 4$\% relative resolution)
for all accepted particles. 
These measurements  yield information on particle mass  
which results in large acceptance (limited resolution)
particle identification.
Four Time--of--Flight walls 
($\approx $ 60 ps resolution)  complement the particle identification
capabilities of the NA49 detector.

A typical resolution in reconstruction of the distance
between collision point and secondary vertex of a two body decay
near the target
is of the order of 1 cm and  is mainly due to the long extrapolation length
needed from the first TPC detector.

\subsection{Statistical Error on  
$\langle D^0 + \overline{D}^0 \rangle_{PbPb}$
in NA49}

In order to estimate the statistical resolution of
a $\langle D^0 + \overline{D}^0 \rangle_{PbPb}$ measurement in NA49
we performed a simulation of signal and background pairs
as expected in this experiment for central
Pb+Pb collisions at 158 A$\cdot$GeV.
In the simulation the standard geometry and magnetic field of
NA49 are assumed \cite{NA49}.
We define the geometrical accpetance of the NA49 TPC's
by the requirement that the  particle trajectories cross more than 20
TPC padrows, at least one of which has to be  in a TPC located
inside the magnetic field.
A parametrization of the momentum resolution as measured by NA49
is included in the simulation.
Concerning particle identification we consider two extreme
cases.
In the first one we assume that no information on particle mass
is available, whereas ideal particle identification is
assumed in the second case.
For the background calculation spectra of charged hadrons
produced in central Pb+Pb collisions at 158 A$\cdot$GeV
as measured by NA49 \cite{PRL} are parametrized in rapidity  ($y$)
and transverse momentum ($p_T$). 
The signal simulation is done assuming a Gaussian ($\sigma = 0.6$)
rapidity distribution and an exponential spectrum in
transverse mass ($T = 300$ MeV) for both $D^0$ and
$\overline{D}^0$ mesons.
The total multiplicity of $D^0$ and $\overline{D}^0$ mesons of 6
as  predicted by
the statistical model (see point 3.2.{\bf D}) is assumed (if needed).
In Fig. 3 we show the $y-p_T$ distribution of $D^0$ mesons 
for which both decay products are in the geometrical acceptance of
the NA49 TPC's.

Before further presentation of results of the simulation we note that
conditions 1 and 2  (see Section 4.1),
which require an accurate measurement of the decay vertex,
can not be used in the exisiting NA49 set--up
to distinguish signal from background pairs.
This is due to the poor resolution in the 
reconstruction of the secondary vertex
from two body decays  of the order of 1 cm, which is 
much larger than the typical flight path of  $D^0$ and $\overline{D}^0$
of about 1 mm.
Therefore background rejection has to be done using conditions 
3,  energy--momentum conservation, and 4, 
isotropy of the decay, only.
We quantify these conditions by studing $M(K^-, \pi^+)$
(or $M(K^+, \pi^-)$) distribution and cos$\Theta$
distribution, where $\Theta$ is the  angle between 
$D$ and $\pi$ meson directions calculated in $D$--meson c.m. system.

The results of the simulation shown below are obtained for the $D^0$ decay,
without using information on particle
mass.
An improvement of the statistical resolution of charm
measurement due to the addition of $\overline{D}^0$ decays and
particle mass information is considered at the end of this section.

In Fig. 4 we show an invariant mass distribution $M(K^-,\pi^+)$
for background pairs obtained using the geometrical acceptance of
NA49.
It is seen that the  $D^0$ meson mass is located in the region of
the monotonically decreasing tail of the distribution, 
well beyond the position
of its maximum.
The $M(K^-,\pi^+)$ distributions for signal and background pairs
in the  interval of $M$ around $m_{D^0}$ are shown in Figs. 5a and
5c, respectively.
For pairs from the interval $\Delta M$ = 4 MeV/c$^2$
around $m_{D^0}$ we plot also the cos$\Theta$ distributions
in Figs. 5b and 5d.

We observe  that in the case of $D^0$
measurement by NA49 we are in the poor resolution limit
i.e. the statistical resolution of signal measurement 
is determined by the number  of background pairs and
is independent of the signal multiplicity. 
It is also clear that rejection of pairs with high 
cos$\Theta$ values should result in an improvement
of statistical significance of the signal measurement. 
We estimate that the maximum significance can be achieved by
accepting pairs with cos$\Theta < 0.7$ and
inside an interval of $\Delta M = $ 4 MeV/c$^2$ (background cuts).
The mean multiplicity of background pairs for this
selection is   
\begin{equation}
\langle N_B \rangle_{NOID} \approx 80,
\end{equation}
where the subscript $_{NOID}$ is used to underline that the number is
obtained without using information on particle identification.
This yields an estimate of the statistical
resolution of $\langle N^E_S \rangle$
\begin{equation}
\sigma_{NOID}(\langle N^E_S \rangle) =
\frac { \sqrt{\langle N_B \rangle}_{NOID} } { \sqrt{N_{EV} } } \approx
\frac { 9 } { \sqrt{N_{EV} } }.
\end{equation}

The resulting correction factors
needed to obtain $\langle D^0 \rangle$ from the estimated signal multiplicity
$\langle N^E_S \rangle$ in the  acceptance are:
\begin{itemize}
\item
$w_{BR}$ = 26.0, for the branching ratio,
\item
$w_{GA}$ = 2.4,  for the geometrical accepatnce,
\item
$w_{BR}$ = 2.3,  for the background cuts.
\end{itemize}
Therefore the statistical resolution of $\langle D^0 \rangle$
can be estimated as:
\begin{equation}
\sigma_{NOID}(\langle D^0 \rangle) = w_{BR} \cdot  w_{GA} \cdot w_{BC} \cdot
\frac { \sqrt{\langle N_B \rangle_{NOID}} } { \sqrt{N_{EV} } } \approx
\frac { 1300 } { \sqrt{N_{EV} } }.
\end{equation}
In the case of ideal particle identification background multiplicity
can be reduced by about a factor of 10, which results in
$\sigma_{ID}(\langle D^0 \rangle) \approx 400/\sqrt{N_{EV} }$.
Therefore
for $10^6$ central Pb+Pb collisions at 158 A$\cdot$GeV 
$\sigma(\langle D^0 \rangle) \approx$ 0.4--1.3 in comparison to
$\langle D^0 \rangle \approx 3$ expected in the case of 
the statistical model.
Statistical significance of measurement of 
$\langle D^0 + \overline{D}^0 \rangle$ is approximately by a factor
of $\sqrt{2}$ better than significance of separate measurements for
$D^0$ or $\overline{D}^0$.

In Fig. 6 we plot $\sigma_{NOID}(\langle D^0 + \overline{D}^0  \rangle)$
and  $\sigma_{ID}(\langle D^0 + \overline{D}^0  \rangle)$ as a 
function of the number of central Pb+Pb collisions at 158 A$\cdot$GeV
registered by NA49.
The different predictions concerning 
$\langle D^0 + \overline{D}^0  \rangle$ discussed in Section 3
are also indicated in Fig. 6 for comparison.

Finally in Fig. 7 we show $M(K^-, \pi^+)$ distributions obtained
using the cos$\Theta$ cut for signal and background pairs for
100 central Pb+Pb collisions.
In this case the number of generated $D^0$ decays was scaled up by a
factor of 100 and therefore the statistical significance of the
observed signal peak corresponds to the significance 
expected for $10^6$ events.

\subsection{Discussion}

The experiment NA49 registered up to now about 10$^6$
central Pb+Pb collisions at 158 A$\cdot$GeV.
As follows from the results of  simulation  presented in Fig. 5
already the analysis of this data should yield significant results
concerning open charm production in Pb+Pb collisions.
This encouraging conclusion is reached mainly due to
three factors:
\begin{itemize}
\item
there are expectations of a significant enhancement of
open charm production in A+A collisions,
\item
a significant fraction of produced $D$--mesons is covered
by the large geometrical acceptance of NA49,
\item
the  good momentum resolution of NA49 allows for a significant reduction of
background even without reconstructing the $D$--meson decay vertex.
\end{itemize}

The results presented in this paper are obtained by a  simple
analysis of simulated data.
This scheme was selected in order to
underline the main concepts and to build intuition concerning
basic ingredients of the problem of $D$--meson measurement
in A+A collisions.
The use of more sophisticated statistical methods
of data analysis (see Ref. \cite{Ga:92}), which e.g. include 
explicitly  the statistical errors of the measured  momentum vector
by using  a kinematical fit, may lead to an improvement of 
the achieved resolution. 

We studied also the statistical resolution of open charm measurement
using three and four body decays of neutral and charged $D$--mesons.
We found that due to high combinatorial background the 
statistical resolution is
much lower than in the case of the two body decay channel.

\newpage

\section{Summary and Conclusions}

The main results presented in this paper can be summarized as follows.
\begin{itemize}
\item
The measurement of open charm production in A+A collisions
gives a unique possibility to test predictions of thermodynamical
and pQCD--based approaches in the region where both are formally
applicable.
\item
Various approaches used to estimate the $D$--meson multiplicity
in central Pb+Pb collisions at 158 A$\cdot$GeV give predictions
which differ by more than a factor 100.
\item
Experimental data already registered by NA49 should allow to obtain
a significant result on open charm production in Pb+Pb collisions.
The analysis can yield the first observation of open charm
signal or it will lead to an estimate of the upper limit
of open charm multiplicity which should significantly narrow the range
of allowed models. 
\end{itemize}

It is obvious that a significant increase of statistics of A+A collisions
at maximum SPS energy registered by NA49 is required for 
continuation of open charm program.
This can be achieved during the already scheduled heavy ion run 
at the CERN SPS  
in the year 2000 and the possible runs beyond.
It is also clear that for a precise measurement of open charm
production an upgrade of the NA49 set--up by a vertex detector, which
allows for an accurate reconstruction of $D$--meson decay vertices,
is needed.

\vspace{0.5cm}

{\bf Acknowledgements}

We thank H. G. Fischer, L. L. Frankfurt, M. I. Gorenstein,
P. Seyboth, R. Stock, H. Str\"obele and G. Roland for
numerous discussions on the subjects coverd in this paper
and comments to the manuscript.

\newpage

{\bf Table 1}
\noindent
The  results on mean multiplicity of 
$D^0 + \overline{D}^0$
mesons produced in p+p and p+N  interactions.
The multiplicty for p+N interactions was estimated
from the data on p+A interactions (used reactions are
listed in the third column) assuming that cross section
for $D$--meson production is proportional to $A$.
For detailed discussion of the data see  text.

\vspace{0.5cm}

\begin{tabular}{|c|c|c|c|c|}
\hline
       &        &         &        &           \cr 
 $p_{LAB}$ [GeV]  &  $\sqrt{s}$ [GeV]  &  Reactions   &~~
$ \langle  D^0 + \overline{D}^0 \rangle \cdot 10^4$
~~&~~
 Reference
~~
\cr
   &      &           &           &      \cr 
\hline
\hline
 200   &  19.4 & p+Si            & 0.47 $\pm$ 0.23  & \cite{Be:88}  \cr
 250   &  21.7 & p+(Be,Al,Cu,W)  & 3.6  $\pm$ 0.9   & \cite{Al:96}  \cr
 360   &  26.0 & p+p             & 6.2  $\pm$ 1.9   & \cite{Ag:84}  \cr
 400   &  27.4 & p+p             & 5.6  $\pm$ 0.8   & \cite{Ag:88}  \cr
 800   &  38.8 & p+p             & 6.5  $\pm$ 2.4   & \cite{Am:88}  \cr
 800   &  38.8 & p+(Be,Au)       & 5.2  $\pm$ 1.0   & \cite{Am:88}  \cr
\hline
\end{tabular}

\newpage

\begin{figure}[p]
\epsfig{file=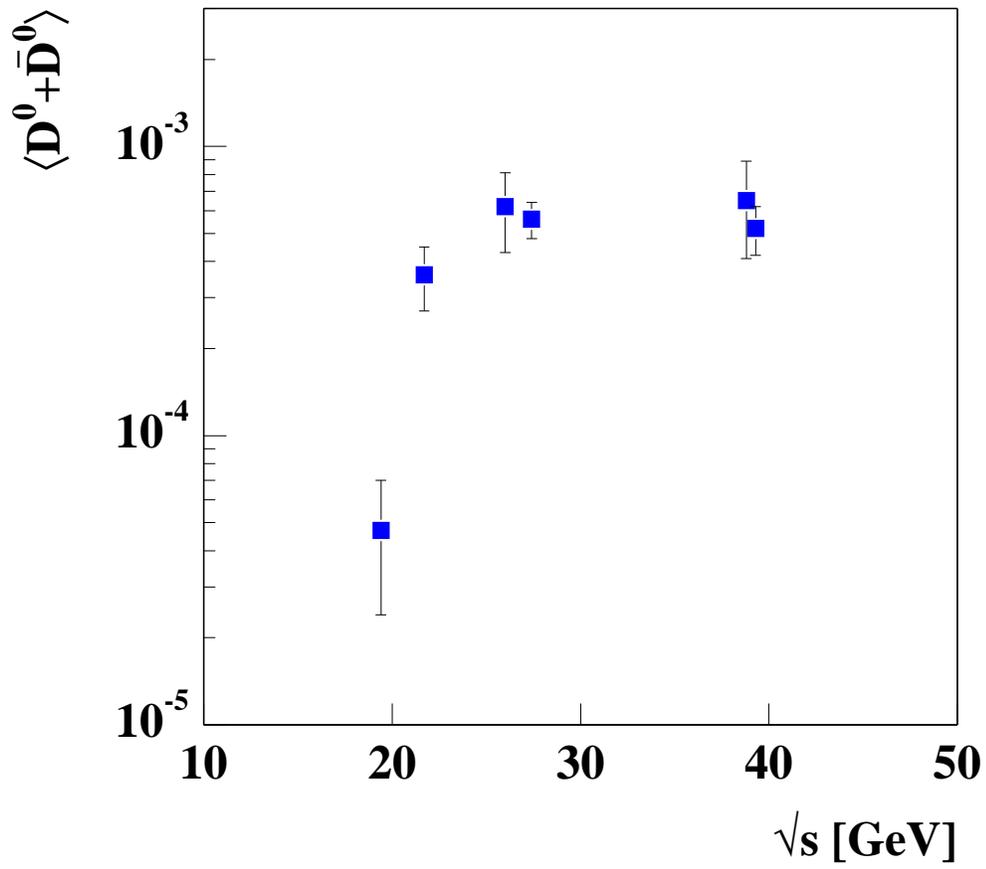,width=14cm}
\caption{
The mean multiplicity of $D^0 + \overline{D}^0$
mesons in proton--nucleon interactions as a function of
collision energy in the c.m. system.
}
\label{fig1}
\end{figure}

\newpage

\begin{figure}[p]
\epsfig{file=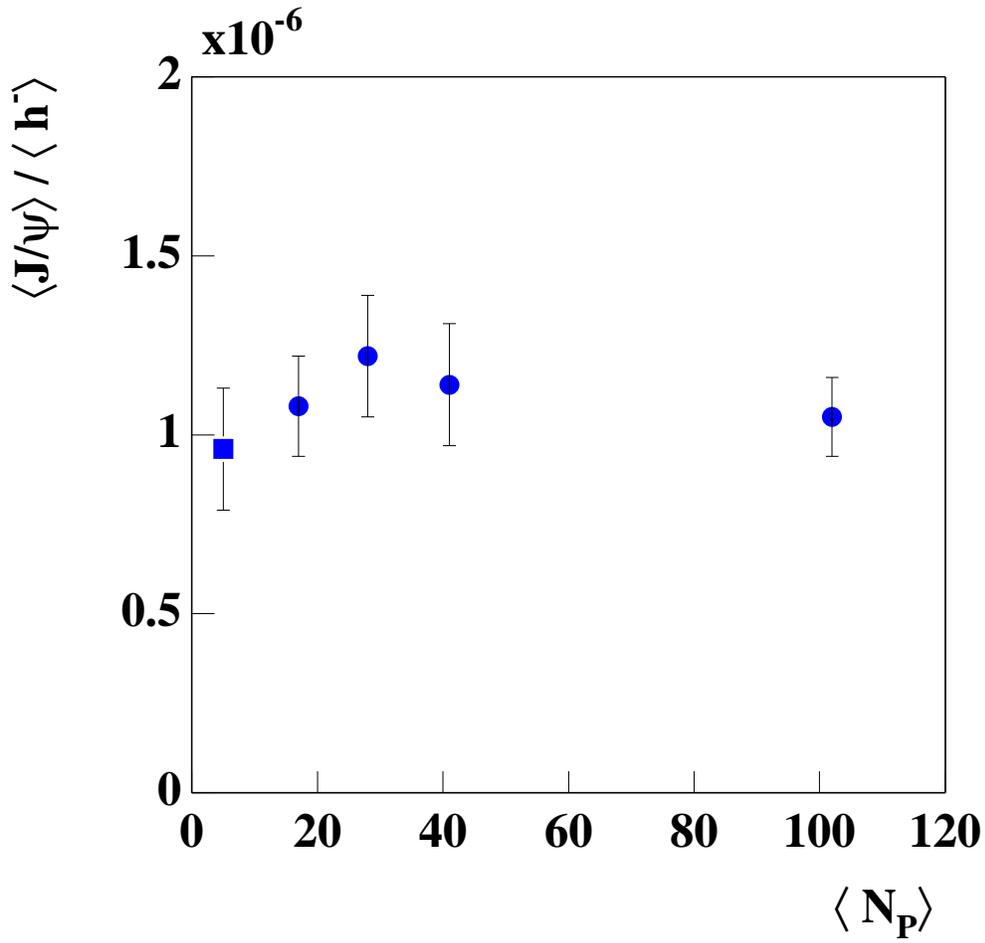,width=14cm}
\caption{
The ratio of the mean multiplicities of $J/\psi$ mesons
and negatively charged hadrons for inelastic nucleon--nucleon (square) and
inelastic O+Cu, O+U, S+U and Pb+Pb (circles) interactions at   
158 A$\cdot$GeV plotted as a function of the mean
number of participant nucleons.
For clarity the N+N point is shifted from
$\langle N_P \rangle = 2$ to $\langle N_P \rangle = 5$.
}
\label{fig2}
\end{figure}

\newpage

\begin{figure}[p]
\epsfig{file=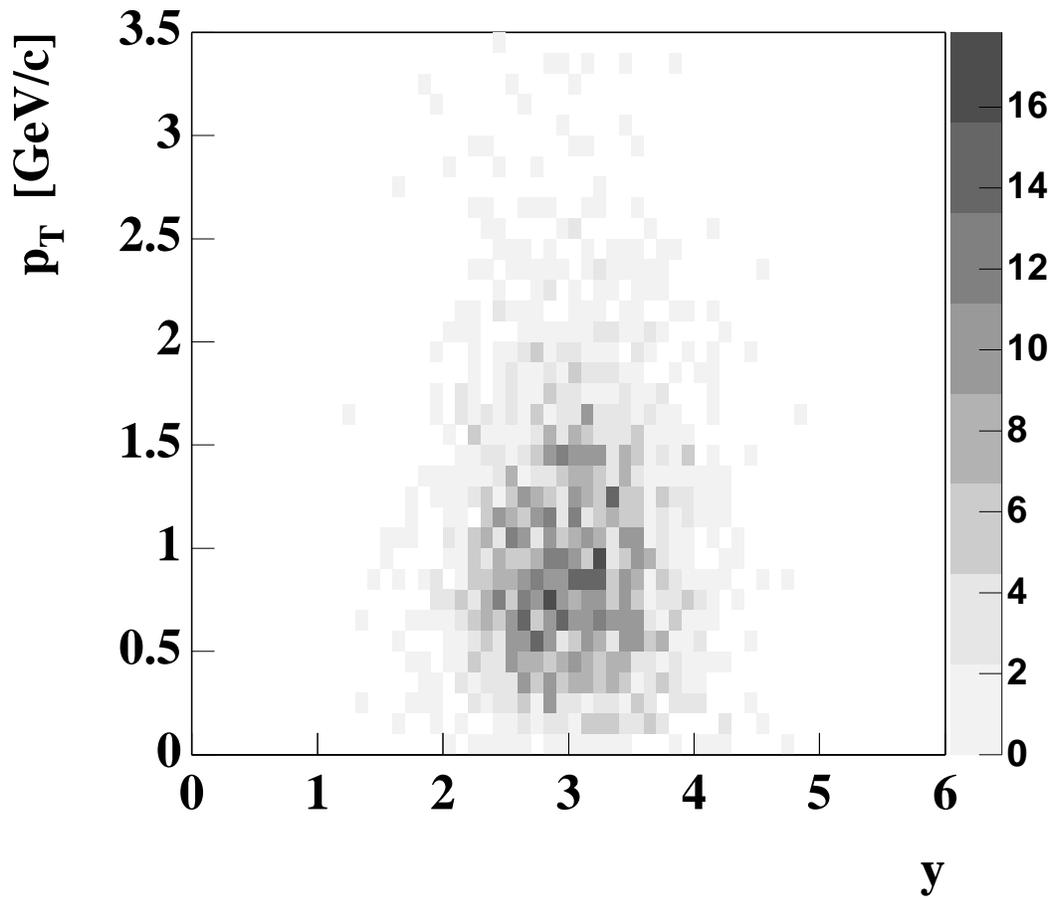,width=14cm}
\caption{
The simulated distribution in transverse momentum and rapidity
of $D^0$ mesons for which both decay products
($ D^0 \rightarrow K^- + \pi^+ $) are reconstructed in
the NA49 TPC's.
The distribution is given in arbitrary units.
}
\label{fig3}
\end{figure}

\newpage

\begin{figure}[p]
\epsfig{file=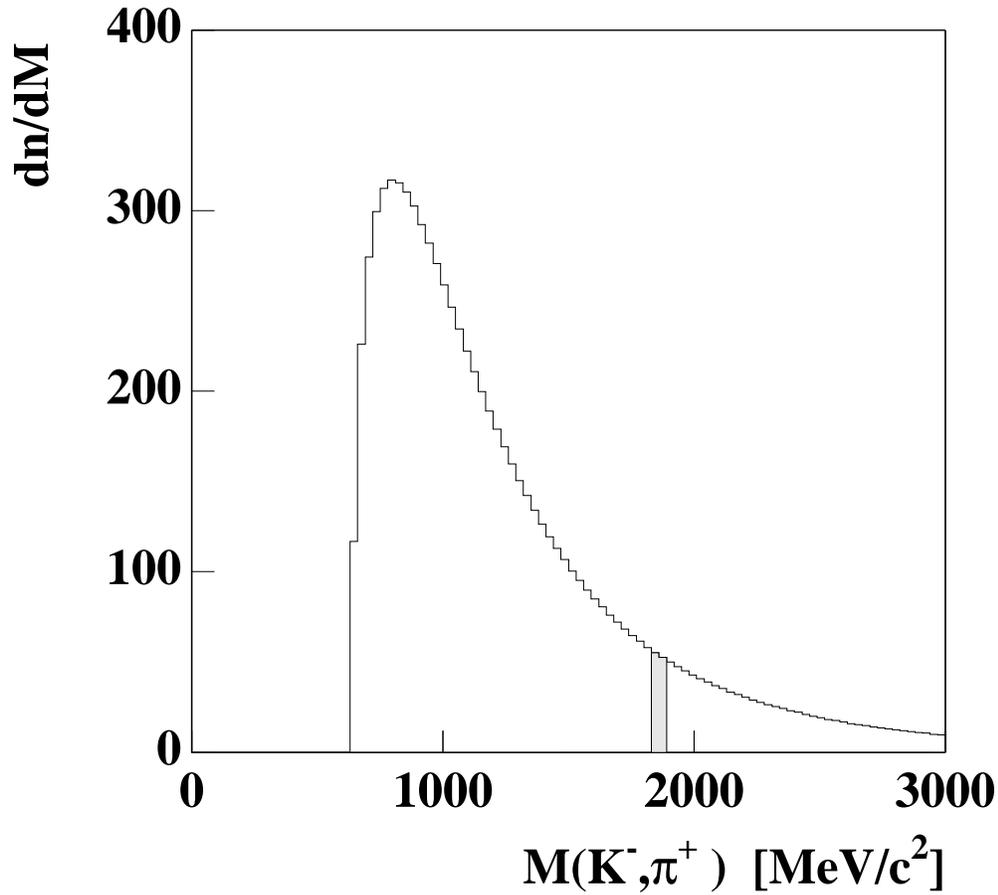,width=14cm}
\caption{
The simulated invariant mass distribution for all combinations
of positively and negatively charged hadrons accepted by
the NA49 TPC's. Masses of positively and negatively charged hadrons
are assumed to be kaon and pion masses, respectively.
The calculation is performed for central Pb+Pb collisions
at 158 A$\cdot$GeV.
The spectrum is normalized per event and it is in (MeV/c$^2$)$^{-1}$
units.  
The shadowed area indicates the invariant mass region around $D^0$
mass further used for detailed analysis.
}
\label{fig4}
\end{figure}

\newpage

\begin{figure}[p]
\epsfig{file=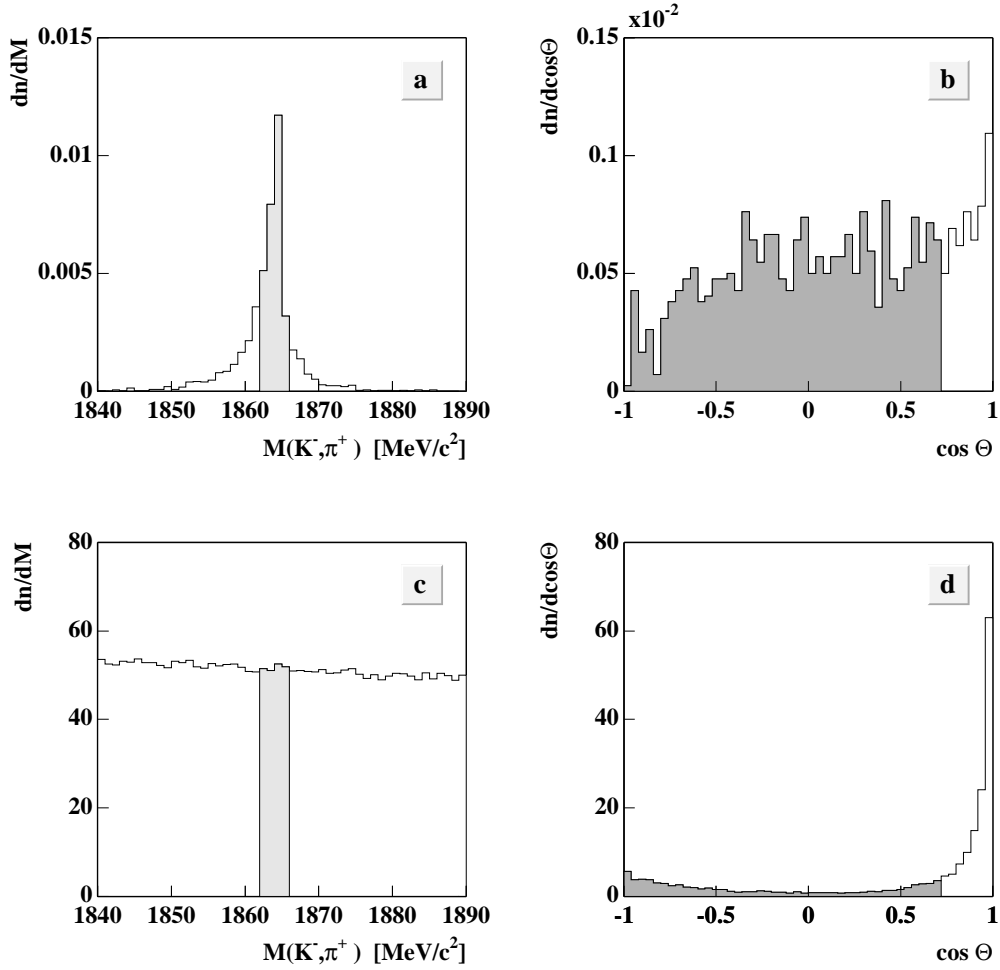,width=14cm}
\caption{
The invariant mass and cos$\Theta$ distributions for
$D^0$ decays (plots {\bf a} and  {\bf b}) and
background (plots {\bf c} and  {\bf d})  pairs.
The cos$\Theta$ distributions are plotted for pairs
from the interval $\Delta M$ = 4 MeV/c$^2$ around 
$m_{D^0}$.
Shadowed areas indicate regions selected for final analysis
(see text for details).
The spectra are normalized per event, the invariant mass
spectra are in (MeV/c$^2$)$^{-1}$ units.
}
\label{fig5}
\end{figure}

\newpage

\begin{figure}[p]
\epsfig{file=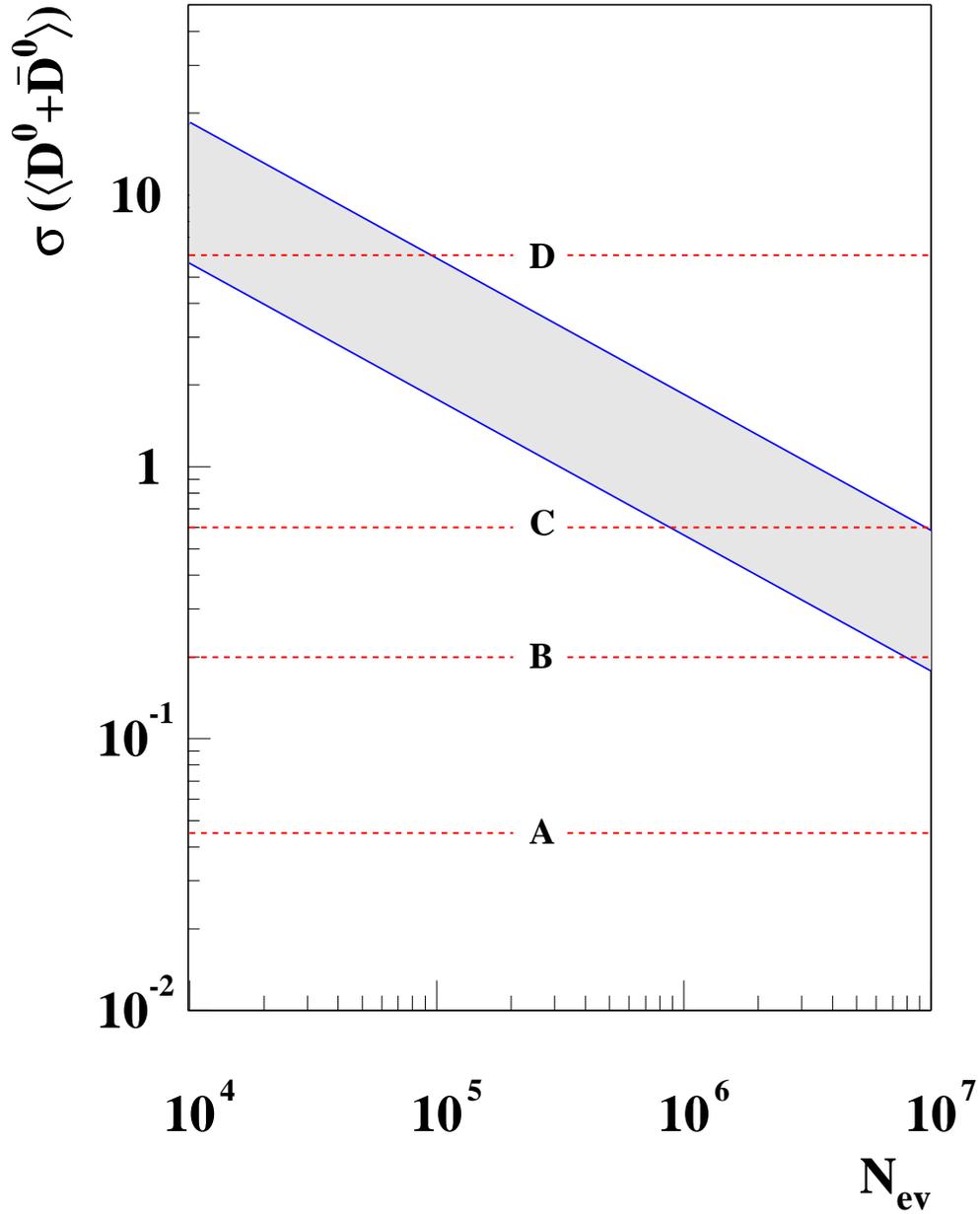,width=14cm}
\caption{
The statistical resolution of the measurement of mean
multiplicity of  $D^0 + \overline{D}^0$ mesons in central
Pb+Pb collisions at 158 A$\cdot$GeV as a function of the
number of analyzed events.
The calculation is performed for the current NA49 set--up assuming
no information on particle mass (upper solid line)
and an ideal particle identification (lower solid line).
The mean multiplicity of $D^0 + \overline{D}^0$ mesons
estimated in four different approaches ({\bf A--D}, see Section 3)
is indicated by dashed lines.
}
\label{fig6}
\end{figure}

\newpage

\begin{figure}[p]
\epsfig{file=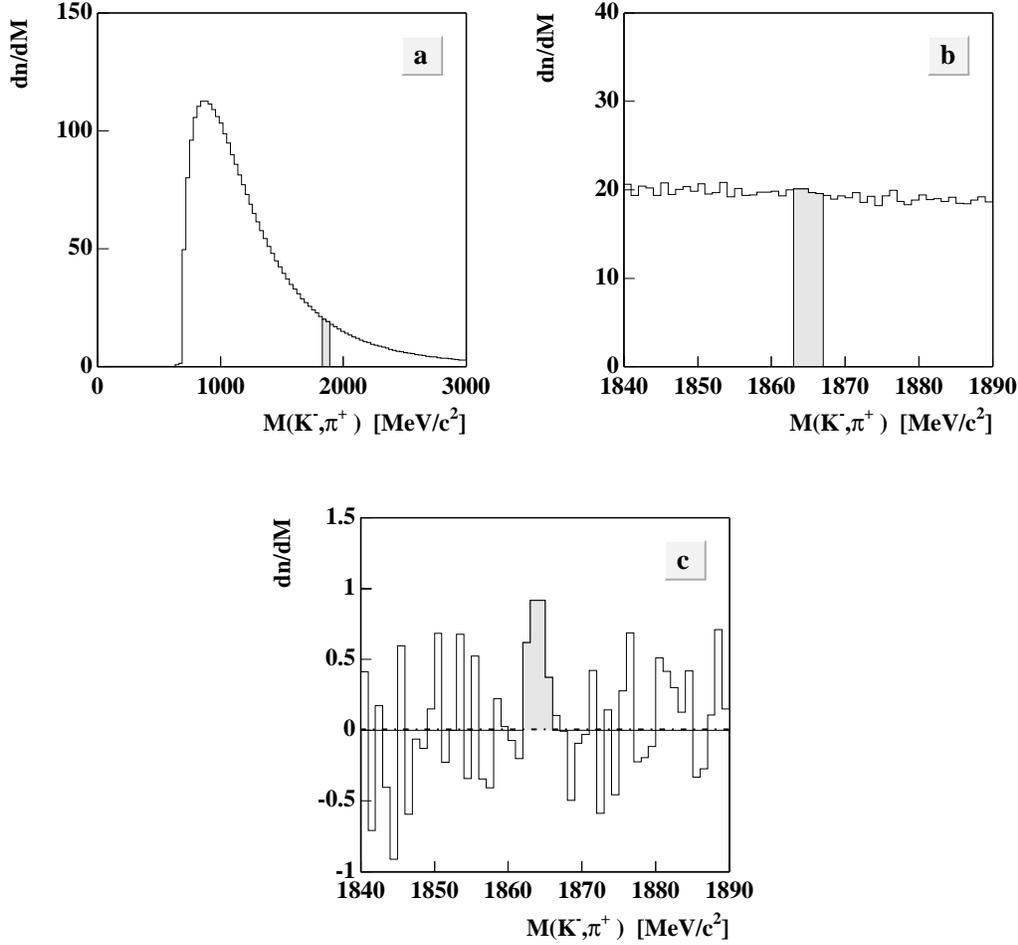,width=14cm}
\caption{
The invariant mass distribution for a sum of signal
($D^0$ decays) and background pairs simulated for
100 central Pb+Pb collisions at 158 A$\cdot$GeV
(plots {\bf a} and {\bf b}).
In the simulation the current NA49 set--up was used and
the number of signal pairs expected in the statistical model
(see 3.2.{\bf D})  was increased by a factor of 100. 
Thus the statistical significance of the  result
corresponds to the significance expected for 10$^6$
central Pb+Pb collisions.   
The difference of the distribution for a sum of signal and background pairs
(shown in plot {\bf b}) and the distribution for background pairs
calculated within the `mixed event' model of background
is shown in plot {\bf c}. 
For all plots only pairs with cos$\Theta < 0.7$ are selected.
All spectra are normalized per event and they are given in 
(MeV/c$^2$)$^{-1}$ units. 
}
\label{fig7}
\end{figure}

\end{document}